\pdfoutput=1

\documentclass[twocolumn,twoside,nofootinbib,showpacs,prd,aps,tightenlines]{revtex4-1}

\usepackage{bm}
\usepackage{amsmath}
\usepackage{graphicx}
\usepackage[hyperfootnotes=true]{hyperref}
\usepackage{xspace}

\newcommand{\eq}[1]{Eq.~\eqref{eq:#1}}
\newcommand{\eqs}[2]{Eqs.~\eqref{eq:#1} and \eqref{eq:#2}}

\newcommand{\subsec}[1]{Sec.~\ref{subsec:#1}}

\newcommand{\app}[1]{Appendix~\ref{app:#1}}
\newcommand{\fig}[1]{Fig.~\ref{fig:#1}}

\def\dpdf#1{F_{#1}}
\def\spdf#1{f_{#1}}
\def\dpdfa#1{{#1}}
\def\spdfa#1{{#1}}
\def\dpdfk#1{{\mathcal F}_{#1}}

\newcommand{\ord}[1]{{\mathcal O}(#1)}

\newcommand{\Mae}[3]{\bigl\langle#1\bigr\rvert#2\bigr\rvert#3\bigr\rangle}
\newcommand{\MAe}[3]{\Bigl\langle#1\Bigr\rvert#2\Bigr\rvert#3\Bigr\rangle}

\newcommand{\ket}[1]{\lvert#1\rangle}

\newcommand{\braket}[2]{\langle#1\rvert#2\rangle}

\newcommand{\nn}{\nonumber}

\newcommand{\df}{\mathrm{d}}
\newcommand{\img}{\mathrm{i}}
\newcommand{\sdt}{\!\cdot\!}

\newcommand{\ga}{\gamma}

\newcommand{\de}{\delta}
\newcommand{\De}{\Delta}
\newcommand{\eps}{\epsilon}

\newcommand{\si}{\sigma}

\newcommand{\W}{\Omega}

\newcommand{\nslash}{n\!\!\!\slash}
\newcommand{\bnslash}{\bar{n}\!\!\!\slash}

\newcommand{\kslash}{k\!\!\!\slash}

\newcommand{\zslash}{z\!\!\!\slash}

\newcommand{\bn}{\bar{n}}

\newcommand{\up}{\uparrow}
\newcommand{\down}{\downarrow}

\newcommand{\lqcd}{\Lambda_\mathrm{QCD}}

\allowdisplaybreaks[2]

\begin{document}


\title{Double Parton Correlations in the Bag Model}

\author{Hsi-Ming Chang}
\author{Aneesh V.~Manohar}
\author{Wouter J.~Waalewijn}

\affiliation{Department of Physics, University of California at San Diego, 9500 Gilman Drive,
La Jolla, CA 92093, U.S.A. \vspace{-0.5ex}}

\begin{abstract}
Double parton scattering is sensitive to correlations between the two partons in the hadron, including correlations in flavor, spin, color, momentum fractions and transverse separation. We obtain a first estimate of the size of these correlations by calculating the corresponding double parton distribution functions in a bag model of the proton. We find significant correlations between  momentum fractions, spin and flavor, but negligible correlations with  transverse separation. The model estimates of the relative importance of these correlations will help experimental studies disentangle them. \\ 

\end{abstract}

\maketitle

\section{Introduction}

High-energy scattering processes such as Drell-Yan production, $p p \to \ell^+\ell^-$, are described by the scattering of two incoming partons, and the cross section is given by the convolution of a partonic scattering cross section $\hat \sigma$ and parton distribution functions (PDFs). Sometimes two hard partonic collisions take place within a single hadronic collision, a process which is known as double parton scattering (DPS). DPS is higher twist, i.e.~it is suppressed by a power of $\lqcd^2/Q^2$, where $Q$ is the partonic center-of-mass energy of the collision. Despite this power suppression, the DPS scattering rate is still large enough that it has become a background for new physics searches at the LHC. For example, DPS contributes to same-sign $WW$ and same-sign dilepton production~\cite{Kulesza:1999zh,Cattaruzza:2005nu,Maina:2009sj,Gaunt:2010pi} and is a background for Higgs studies in the channel $pp \to WH \to \ell \nu b\bar b$ \cite{DelFabbro:1999tf,Hussein:2006xr,Bandurin:2010gn,Berger:2011ep}. DPS has been observed at the LHC;  a preliminary study using $33\, \text{pb}^{-1}$ of data found that 16\% of the $W+2$ jet events were due to DPS~\cite{ATLAS-CONF-2011-160}.

In the original work on DPS, the cross section was written as~\cite{Paver:1982yp}
\begin{align}  \label{eq:si_old}
  \df \si &= \frac{1}{S} \sum_{i,j,k,l} \int\! \df^2 \mathbf{z_\perp}\, \dpdf{ij}(x_1,x_2,\mathbf{z_\perp},\mu) \dpdf{kl}(x_3,x_4,\mathbf{z_\perp},\mu)  \nn \\ & \quad \times 
  \hat \si_{ik}(x_1 x_3 \sqrt{s},\mu) \hat \si_{jl}(x_2 x_4 \sqrt{s},\mu)
\,.\end{align}
The double parton distribution function (dPDF) $\dpdf{ij}(x_1,x_2,\mathbf{z_\perp})$ describes the probability of finding two partons with flavors $i,j=g,u, \bar u, d, \dots$, longitudinal momentum fractions $x_1,x_2$ and transverse separation $\mathbf{z_\perp}$ inside the hadron. The partonic cross sections $\hat \si$ describe the short-distance processes, and $S$ is a symmetry factor that arises for identical particles in the final state. \eq{si_old} ignores additional contributions that are sensitive to diparton correlations in flavor, spin and color, as well as parton-exchange interference contributions~\cite{Mekhfi:1985dv,Diehl:2011tt,Diehl:2011yj,Manohar:2012jr}. These correlations are present in QCD, and  one of our goals is to estimate the size of these effects.

It is commonly assumed in DPS studies that the dependence on the transverse separation is uncorrelated with the momentum fractions or parton flavors, 
\begin{equation} \label{eq:zfact}
  \dpdf{ij}(x_1,x_2,\mathbf{z_\perp},\mu) = \dpdf{ij}(x_1,x_2,\mu) G(\mathbf{z_\perp},\mu)
\,.\end{equation}
In addition, a factorized ansatz is often made,
\begin{equation} \label{eq:xfact}
\begin{split}
  &\dpdf{ij}(x_1,x_2,\mu) \\
   &= \spdf{i}(x_1,\mu) \spdf{j}(x_2,\mu)\, \theta(1-x_1-x_2) (1-x_1-x_2)^n
\,,
\end{split}
\end{equation}
where $f$ denotes the usual (single) PDF. The factor $\theta(1-x_1-x_2) (1-x_1-x_2)^n$ smoothly imposes the kinematic constraint $x_1+x_2 \leq 1$, and different values of the parameter $n>0$ have been considered. 

The dPDF is a nonperturbative function, but once it is known at a certain scale $\mu$, its renormalization group evolution can be used to evaluate it at a different scale. The evolution of $\dpdf{ij}(x_1,x_2,\mu)$ was determined a long time ago~\cite{Kirschner:1979im,Shelest:1982dg}. It has recently been extended to include the $\mathbf{z_\perp}$ dependence and describe correlation and interference dPDFs~\cite{Diehl:2011tt,Diehl:2011yj,Manohar:2012jr,Manohar:DPS2}. The color-correlated and interference dPDFs are all Sudakov suppressed at high energies~\cite{Mekhfi:1988kj,Manohar:2012jr} and will, therefore, not be considered. 

Eventually the dPDFs will be determined by fitting to experimental data, just as for the usual PDFs. Reference~\cite{Kasemets:2012pr} goes a step in this direction, showing how angular correlations in double Drell-Yan production may be used to study spin correlations in dPDFs.  In this paper, we determine the dPDFs at a low scale $\mu \sim \lqcd$ using a bag model for the proton~\cite{Chodos:1974pn}. This model calculation provides an estimate of the importance of various diparton correlations, which can be used to guide the experimental analysis. It  also provides an estimate of  dPDF distributions in the absence of more accurate determinations directly from experiment.

We follow some of the existing structure function calculations in the bag model~\cite{Jaffe:1974nj,Benesh:1987ie,Schreiber:1991tc}. There are obvious limitations to this approach, just as for bag model calculations of the usual PDFs. First of all, the bag model only describes valence quarks. Bag model calculations are only meaningful when the fields in the dPDF act inside the bag, which restricts the momentum fractions $x \gtrsim 1/(2MR)\sim 0.1$, where $M$ is the proton mass and $R$ is the bag radius. Finally, the bag was treated as rigid in the early literature, Ref.~\cite{Jaffe:1974nj}. A consequence is that momentum is not conserved,  and parton distributions do not vanish outside the physical region ($x>1$). Alternative treatments of the bag were proposed to alleviate this problem~\cite{Benesh:1987ie,Schreiber:1991tc,Wang:1982tz}. We emphasize that we are not attempting to use the most sophisticated bag model description of the proton. Rather, we simply want to provide a first estimate of the size of the various correlation effects. Bag model PDFs are usually chosen as the initial value of PDFs at a low scale $\mu \sim \lqcd$, which are then evolved to higher scales using their QCD evolution. Since in the bag model the valence quarks carry all the momentum, this initial scale $\mu$ needs to be taken quite low~\cite{Benesh:1987ie}.

We also  investigate \eqs{zfact}{xfact} in this paper, using our bag model results. We find that \eq{zfact} holds reasonably well, but \eq{xfact} is badly violated. Problems with \eq{xfact} have already been pointed out in Ref.~\cite{Gaunt:2009re}, using sum rules and the evolution of the dPDF. (Though \eq{xfact} may still be approximately true when one of the momentum fractions $x_i$ is small; see, e.g., Ref.~\cite{Snigirev:2010ds}.) In the simplest bag models of the type we consider, the color-correlated dPDFs $F^T$ are given by $-2/3$ times the color-direct dPDFs $F^1$, since diquarks are in a $\mathbf{\overline 3}$ representation of color.

\section{Calculation}
\label{sec:calc}

We briefly summarize the ingredients of the bag model~\cite{Chodos:1974pn} that are needed to calculate the dPDFs. The bag model wave functions are the solutions of the massless Dirac equation in a spherical cavity of radius $R$. We only need the ground state, which is given by 
\begin{align} \label{eq:psi_bag}
  \Psi_m({\bf r},t) = N \begin{pmatrix}
  j_{0}(\W |{\bf r}|/R) \ \chi_m \\[5pt]
  \img\,  {\bf \hat r} \sdt {\bm \si}\, j_1(\W |{\bf r}|/R)\ \chi_m
  \end{pmatrix}
  e^{-\img \W t/R}
\,,\end{align}
for a bag centered at the origin. Here $\W=ER$, with $E$ the energy of the particle,
\begin{equation}
  \!\! N^2 = \frac{1}{R^3} \frac{\W^4}{\W^2 - \sin^2 \W}
  \,, \quad
  \chi_m = \frac{1}{\sqrt{4\pi}} \begin{pmatrix} \de_{m,\up} \\ \de_{m,\down} \end{pmatrix}  \label{eq:chidef}
\,,\end{equation}
$j_i$ are  spherical Bessel functions, and $m=\uparrow,\downarrow$. The condition that the color current does not flow through the boundary $r^\mu \Psi \ga_\mu T^A \Psi |_{|{\bf r}|=R} = 0$ leads to
\begin{align} 
  j_0(\W) &= j_1(\W)
  \quad \Rightarrow \quad
  \Omega \approx 2.043
\,,\end{align}
and we will take $R=1.6$ fm in our numerical analysis.

The quark field is expanded in terms of bag wave functions,  quark creation and annihilation operators $a_i(\mathbf{a})$, $a_i^\dagger(\mathbf{a})$ and antiquark creation and annihilation operators $b_i(\mathbf{a})$, $b_i^\dagger(\mathbf{a})$. These operators create or annihilate quarks and antiquarks in a bag centered at $\mathbf{r=a}$ [see \eq{field_bag}].

The spin-up proton wave function is given in terms of the standard quark model wave functions as
\begin{align} \label{eq:ketp}
\frac{1}{\sqrt 6}\ket{uud} \left(2 \ket{\up \up \down} -\ket{\up \down \up }- \ket{\down \up \up}\right)
\,.\end{align}
As usual, the color indices are suppressed, and the wave function has to be symmetrized over permutations. Ignoring color, one can also write the wave function in terms of bosonic~\cite{Manohar:2004qw} creation operators,
\begin{align} \label{eq:ketp1}
  \ket{P\up,{\bf r = a}} &= \frac{1}{\sqrt{3}} \Big[a_{u\up}^\dagger({\bf a}) a_{u\up}^\dagger({\bf a}) a_{d\down}^\dagger({\bf a})
 \nn \\  & \quad
 - a_{u\up}^\dagger({\bf a}) a_{u\down}^\dagger({\bf a}) a_{d\up}^\dagger({\bf a})\Big] \ket{0,{\bf r = a}}
\,.\end{align}
Here $\ket{P\up,{\bf r = a}}$ and $\ket{0,{\bf r = a}}$ are the proton and empty bag state, respectively, both at position $\bf a$. The $a_{qm}^\dagger({\bf a})$ denotes the creation operator for a quark of flavor $q$ with spin $m$ in a bag at position $\bf a$. 

An important difference between various calculations in the literature is the treatment of the overlap between empty bags at different positions,
\begin{align} \label{eq:bag_overlap}
\braket{0,{\bf r=a}}{0,{\bf r=b}}&=\de^3({\bf a-b}) & &\text{in Ref.~\cite{Jaffe:1974nj}}
\,,\nn \\
\braket{0,{\bf r=a}}{0,{\bf r=b}}&=1 & &\text{in Refs.~\cite{Benesh:1987ie,Schreiber:1991tc}}
\,.\end{align}
These opposite limits treat the bags as either completely rigid or fully flexible, and the latter will be our default. We will return to the rigid bag in \subsec{jaffe}. To account for the displacement between bags, we follow Ref.~\cite{Benesh:1987ie} in taking
\begin{align} \label{eq:comm}
  \{a_i({\bf a}), a_j^\dagger({\bf b})\} &= \de_{ij}\int\! \df^3 x\, \Psi^\dagger_j({\bf x-b}) \Psi_i({\bf x-a})
\,.\end{align}
For the rigid bag these are replaced by the familiar anticommutation relations
\begin{align}
  \{a_i, a_j^\dagger\} &= \de_{ij}  
\,,\end{align}
where we only need the relation when $a$ and $a^\dagger$ are at the same bag position, because of \eq{bag_overlap}.

The proton state with momentum ${\bf p}$ is constructed using the Peierls-Yoccoz (PY)  projection~\cite{Peierls:1957er},
\begin{align}
  \ket{P,{\bf p}} = \frac{1}{\phi_3({\bf p})} \int\! \df^3 {\bf a}\, e^{\img {\bf a \cdot p}}\, \ket{P,{\bf r = a}}
\,,\end{align}
where $\phi_3({\bf p})$ fixes the (nonrelativistic) normalization of the state. The functions $\phi_n({\bf p})$ are given by
\begin{align}
|\phi_n({\bf p})|^2 
& = \int\! \df {\bf a}\, e^{-\img {\bf p} \cdot {\bf a}} \Big[\int\! \df {\bf x}\,
 \Psi^\dagger({\bf x}-{\bf a}) \Psi({\bf x}) \Big]^n \label{eq:phin}
\,,\end{align}
which we will need for $n=1,2,3$.

The final ingredient  is the expression for quark fields acting in the bag. The field for a $u$ quark relative to a bag at ${\bf a}$ is given by~\cite{Benesh:1987ie}
\begin{align} \label{eq:field_bag}
  u({\bf x},t)= \sum_{m=\up,\down} a_{um}({\bf a}) \Psi_m({\bf x- a}) e^{-\img \W t/R} + \dots
\,.\end{align}
Here ``$\dots$" denotes contributions from other bag states that will not be needed\footnote{We will not consider the so-called z graph or four-quark intermediate state contribution~\cite{Jaffe:1974nj,Schreiber:1991tc}, where the field creates an antiquark. This only contributes at small $x$ and is thus outside the range of validity of the calculation.}. 
The expression for $d$ quarks is similar.

\subsection{Single PDF}

We first summarize the well-known calculation of  the (single) PDF in the bag model. The light-cone vectors are
\begin{align}
n^\mu = (1,0,0,1)\,, \quad
\bn^\mu = (1,0,0,-1)\,,
\end{align}
and we assume the light-cone gauge $n \sdt A = 0$. In the proton rest frame, where $p^\mu = (M,{\bf 0})$,\footnote{We also use the notation $q$ for the PDF $f_q$, and $qq$, $q \Delta q$, $\ldots$ for the dPDFs $F_{qq}$, $F_{q \Delta q}$, $\ldots$.}
\begin{widetext}
\begin{align} \label{eq:fq}
\spdfa{q}(x) &= 2M \int\! \frac{\df z^+}{4\pi}\, e^{-\img x M z^+/2} 
	\MAe{P,{\bf p = 0}}{\bar q \Bigl(z^+\frac{\bn}{2}\Bigr) 
	\frac{\bnslash}{2}\, q(0)}{P,{\bf p = 0}} \nn \\
	&= \sum_{m_1,m_2=\up,\down} \Mae{P,{\bf r = 0}}{a_{qm_1}^\dagger({\bf 0})
	  a_{qm_2}({\bf 0})}{P,{\bf r = 0}} 
	  2M\int\! \frac{\df z^+}{4\pi}\,
	  \frac{\df{\bf k}_1}{(2\pi)^3}\,\frac{\df{\bf k}_2}{(2\pi)^3}\,\frac{\df{\bf k}_3}{(2\pi)^3}\,
	  e^{-\img (xM -\frac{\Omega}{R}+k_{1z})z^+/2} \nn \\ 
	& \quad \times 
  	  (2\pi)^3\delta({\bf k}_1-{\bf k}_3) (2\pi)^3\delta({\bf k}_2-{\bf k}_3)\,
  	  \bar{\widetilde{\Psi}}_{m_1}({\bf k}_1) \frac{\bnslash}{2} 
	  \widetilde{\Psi}_{m_2}({\bf k}_2) \frac{|\phi_2({\bf k}_3)|^2}{|\phi_3(0)|^2} \nn \\
	&= \sum_{m_1,m_2=\up,\down} \Mae{P,{\bf r = 0}}{a_{qm_1}^\dagger({\bf 0})
	  a_{qm_2}({\bf 0})}{P,{\bf r = 0}} \,
	  2M \int\! \frac{d{\bf k}}{(2\pi)^3}\, 
	  \bar{\widetilde{\Psi}}_{m_1}({\bf k}) \frac{\bnslash}{2} 
	  \widetilde{\Psi}_{m_2}({\bf k}) \frac{|\phi_2({\bf k})|^2}{|\phi_3(0)|^2}\,
	  \delta \Big( x M -\frac{\Omega}{R}+k_z \Big) \nn \\
	&= \sum_{m_1,m_2=\up,\down} \Mae{P,{\bf r = 0}}{a_{qm_1}^\dagger({\bf 0})
	  a_{qm_2}({\bf 0})}{P,{\bf r = 0}} 
	  \frac{2M}{(2\pi)^2} 
	  \int_{|\W/R - xM|}^\infty\!\!\!\! \df |{\bf k}|\, |{\bf k}|\, 
	  \bar{\widetilde{\Psi}}_{m}({\bf k}) \frac{\bnslash}{2} 
	  \widetilde{\Psi}_{m}({\bf k}) \frac{|\phi_2({\bf k})|^2}{|\phi_3(0)|^2}	 
\,.\end{align}
\end{widetext}
Here $z^+ = n \cdot z$, $\widetilde{\Psi}$ denotes the Fourier transform of $\Psi$, and $\phi_2$ is given by \eq{phin}. The overall factor of $2M$ is due to the nonrelativistic normalization of states. The delta function on the fourth line sets
\begin{align}
  k_z = \frac{\W}{R} - xM
\,,\end{align}
implying that the peak of the PDF is at $x = \W/(MR)$, independent of the quark flavor. This disagreement with experimental measurements of $\spdfa{u}$ and $\spdfa{d}$ may be alleviated by refining the bag model; see, e.g., Ref.~\cite{Close:1988br}. We will restrict ourselves to the simplest bag models in this paper, so its limitations should be kept in mind while using the results.

In using \eq{field_bag} we assumed that the field $\bar q$ acts at the position of the bag of the left state and $q$ at the position of the bag of the right state~\cite{Benesh:1987ie}. The matrix element of \eq{fq} contains all the dependence on the spin-flavor wave function of \eq{ketp}, which is connected with the spin of the bag wave functions through the sum on $m_{1,2}$. For the unpolarized single PDF only $m_1=m_2$ contributes, and the matrix element simply counts the number of quarks of a given flavor $q$ in the proton,
\begin{figure}[b!]
\centering
\includegraphics[width=0.4\textwidth]{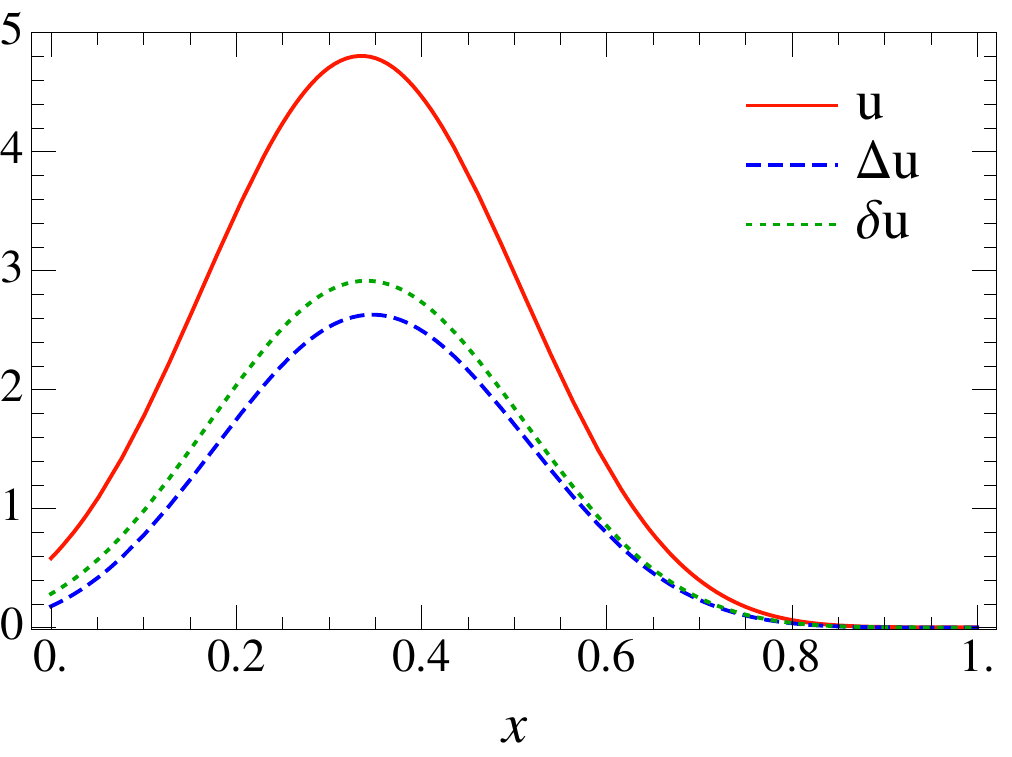}
\caption{The proton  PDFs $\spdfa{u}$ (solid red), $\spdfa{\De u}$ (dashed blue) and $\spdfa{\de u}$ (dotted green). }
\label{fig:sspins}
\end{figure}
%
%
%
\begin{align}
  n_q = \sum_{m=\up,\down} \Mae{P,{\bf r = 0}}{a_{qm}^\dagger({\bf 0}) a_{qm}({\bf 0})}{P,{\bf r = 0}} 
\,.\end{align}
The extension of \eq{fq} to longitudinal and transversely polarized quark distributions is given by replacing $\frac{\bnslash}{2}$ by $\frac{\bnslash}{2} \ga_5$ for $\spdfa{\De q}$ and $\frac{\bnslash}{2} \ga_\perp^\mu \ga_5$ for $\spdfa{\de q}$. $\spdfa{\Delta q}$ and $\spdfa{\delta q}$ only contribute in processes involving longitudinally and transversely polarized protons, respectively. The matrix elements required are evaluated in \subsec{spin_cor}. To aid the evaluation of the remaining integral in \eq{fq}, convenient expressions for the functions $\phi_i$ and the bag wave function in momentum space are given in \app{integrals}. The resulting PDFs are compared in \fig{sspins}.

The spatial distribution of partons inside the nucleon are also probed by the electromagnetic form factors, which are independent of the renormalization scale. They have been calculated within the bag model that we are using, showing reasonable agreement with experiment~\cite{Barnhill:1978ui}. Calculations of form factors for more sophisticated bag models can, for example, be found in Refs.~\cite{Betz:1983dy,Lu:1997sd,Miller:2002ig}.

\subsection{Double PDF}

We calculate the double PDF using the definitions in Ref.~\cite{Manohar:2012jr}. We will not consider color correlated or interference double PDFs, since these are Sudakov suppressed. The spin-averaged dPDF $\dpdf{q_1q_2}(x_1,x_2,{\bf z}_\perp)$ is defined as
\begin{widetext}
\begin{align} \label{eq:dPDF1a}
\dpdf{q_1 q_2}(x_1,x_2,{\bf z}_\perp) 
&= -8\pi M^2 \int \frac{\df z_1^+}{4\pi} \frac{\df z_2^+}{4\pi} \frac{\df z_3^+}{4\pi}  e^{-\img x_1 M  z_1^+/2}\, e^{-\img x_2 M z_2^+/2}\, e^{\img x_1 M z_3^+/2} 
\nn \\ & \quad \times
\MAe{P,{\bf p = 0}}{\Big[\overline q_1 \Big(z_1^+ \frac{\bn}{2}+z_\perp\Big) \frac{\bnslash}{2} \Big]_c
\Big[\overline q_2 \Big(z_2^+ \frac{\bn}{2}\Big) \frac{\bnslash}{2} \Big]_d
 q_{1,c}\Big(z_3^+ \frac{\bn}{2}+z_\perp\Big) q_{2,d}(0)}{P,{\bf p = 0}}
\,.\end{align}
It is convenient to work in terms of the Fourier-transformed distribution $\dpdf{q_1 q_2}(x_1,x_2,{\bf k}_\perp)$. Evaluated in the bag model, 
\begin{align} \label{eq:dPDF1b}
\dpdf{q_1 q_2}(x_1,x_2,{\bf k}_\perp) 
&= \int\! \df^2{\bf z_\perp}\, e^{\img \bf z_\perp \cdot k_\perp} \dpdf{q_1 q_2}(x_1,x_2,{\bf z}_\perp)
 \\ &
= \sum_{m_1,m_2,m_3,m_4} \Mae{P,{\bf r = 0}}{a_{q_1m_1}^\dagger({\bf 0}) a_{q_2 m_2}^\dagger({\bf 0}) a_{q_2 m_4}({\bf 0}) a_{q_1m_3}({\bf 0})}{P,{\bf r = 0}} 
\nn \\ & \quad \times
8\pi\, M^2 \int\!  \frac{\df {\bf k}_1}{(2\pi)^3}\,  \frac{\df {\bf k}_2}{(2\pi)^3}\,  \frac{\df {\bf k}_3}{(2\pi)^3}\,
\de\Big(x_1 M - \frac{\W}{R} + {\bf k}_{1z}\Big)\, \de\Big(x_2 M - \frac{\W}{R} + {\bf k}_{2z}\Big)\, \de\Big(x_1 M - \frac{\W}{R} + {\bf k}_{3z}\Big)  
\nn \\ & \quad \times
(2\pi)^2 \de^2({\bf k}_{1\perp} - {\bf k}_{3\perp} - {\bf k}_\perp) \, \bar{\widetilde{\Psi}}_{m_1}({\bf k}_1) \frac{\bnslash}{2} \widetilde{\Psi}_{m_3}({\bf k}_3)\, \bar{\widetilde{\Psi}}_{m_2}({\bf k}_2) \frac{\bnslash}{2} \widetilde{\Psi}_{m_4}({\bf k}_1 + {\bf k}_2 - {\bf k}_3) \frac{|\phi_1({\bf k}_1 + {\bf k}_2)|^2}{|\phi_3(0)|^2}
\,,\nn\end{align}
\end{widetext}
where $\phi_1$ is given by \eq{phin}. Results for the matrix elements on the second line of \eq{dPDF1b} are given in \subsec{spin_cor}. The remaining integrals were numerically performed using the expressions in \app{integrals} and the CUBA integration package~\cite{Hahn:2004fe}.

\subsection{Spin Correlations}
\label{subsec:spin_cor}

The computation of spin-correlated dPDFs is almost identical to \eq{dPDF1b}. For $F(x_1,x_2,{\bf z}_\perp)$ the spinors $\frac{\bnslash}{2} \otimes \frac{\bnslash}{2}$ in \eq{dPDF1a} are replaced by~\cite{Diehl:2011tt,Diehl:2011yj,Manohar:2012jr}
\begin{align}
\begin{tabular}{|l|c|}
\hline 
 $\dpdf{\De q_1 \De q_2}$ \ & $\frac{\bnslash}{2}\, \ga_5 \otimes \frac{\bnslash}{2}\, \ga_5$ \\
 $\dpdf{\de q_1 \de q_2}$ & $\frac{\bnslash}{2} \ga_\perp^\mu \ga_5 \otimes \frac{\bnslash}{2} \ga^\perp_\mu \ga_5$ \\
 $\dpdf{q_1 \de q_2}$ & $-\frac{1}{M {\bf z}_\perp^2}\, \frac{\bnslash}{2} \otimes \frac{\bnslash}{2}\, \ga_\perp^\mu \eps_{\mu \nu} z_\perp^\nu \ga_5$ \\
 $\dpdf{\De q_1 \de q_2}$ & $-\frac{1}{M {\bf z}_\perp^2}\, \frac{\bnslash}{2}\, \ga_5 \otimes \frac{\bnslash}{2}\, \zslash_\perp \ga_5$ \\
 $\dpdf{\de q_1 \de q_2}^t$ & \ $\frac{2}{M^2|{\bf z}_\perp|^4} (z_\perp^\mu z_\perp^\nu + \frac{1}{2} {\bf z}_\perp^2 g_\perp^{\mu\nu}) \frac{\bnslash}{2} \ga_\mu \ga_5 \otimes \frac{\bnslash}{2} \ga_\nu \ga_5$
 \\[1ex] \hline
 \end{tabular}
\nn\end{align}
As in \eq{dPDF1b}, we switch to momentum space, for which it is convenient to modify some of the spin structures:
\begin{align}
\begin{tabular}{|l|c|}
\hline 
 $\dpdfk{q_1 \de q_2}$ & $\frac{\img M}{{\bf k}_\perp^2}\, \frac{\bnslash}{2} \otimes \frac{\bnslash}{2}\, \ga_\perp^\mu \eps_{\mu \nu} k_\perp^\nu \ga_5$ \\
 $\dpdfk{\De q_1 \de q_2}$ & $\frac{\img M}{{\bf k}_\perp^2}\, \frac{\bnslash}{2}\, \ga_5 \otimes \frac{\bnslash}{2}\, \kslash_\perp \ga_5$ \\
 $\dpdfk{\de q_1 \de q_2}^t$ & \ $\frac{2M^2}{|{\bf k}_\perp|^4} (k_\perp^\mu k_\perp^\nu + \frac{1}{2} {\bf k}_\perp^2 g_\perp^{\mu\nu}) \frac{\bnslash}{2} \ga_\mu \ga_5 \otimes \frac{\bnslash}{2} \ga_\nu \ga_5$
 \\[1ex] \hline
 \end{tabular}
\nn\end{align}
We will always use these momentum-space spin structures in plots. The relationship between $\dpdfk{}$ and $F$ is not simply a Fourier transform, and is given in \app{spins}. The additional factors of $-\img$ in $\dpdfk{q_1 \de q_2}$ and $\dpdfk{\De q_1 \de q_2}$ ensure that these dPDFs are real. The spin structure $\dpdfa{\De q_1 \de q_2}$ vanishes in our calculation. Assuming for simplicity that ${\bf k}$ is along the $x$ direction, this follows from the reflection $k_{1y} \to - k_{1y}$, $k_{2y} \to -k_{2y}$, under which the integrand in \eq{dPDF1b} is odd. Though this is due to the form of the bag model matrix elements, it suggests that the spin structure $\De q_1 \de q_2$ is likely smaller than the others.

We now evaluate the spin-flavor matrix elements that enter in the single and double PDFs. Since we suppressed the \emph{antisymmetric} color wave function of the proton, the creation and annihilation operators essentially satisfy \emph{commutation} relations. 
For the unpolarized and longitudinally polarized single PDF, only $m_1=m_2$ contributes, and we find the weighting:
\begin{align}
\begin{tabular}{|ll|c|}
\hline
 $q$ & $m$ & $\Mae{P \up}{a_{qm}^\dagger a_{qm}}{P \up}$ \\ \hline
 $u$ & $\up$ & 5/3 \\
 $u$ & $\down$ & 1/3 \\
 $d$ & $\up$ & 1/3 \\ 
 $d$ & $\down$ & 2/3 \\ \hline
\end{tabular}
\nn\end{align}
For $\spdfa{\delta q}$ we need a transversely polarized proton
\begin{align}
\ket{P\to} = \frac{1}{\sqrt{2}} (\ket{P\up} + \ket{P\down})
\,.\end{align}
The nonvanishing matrix elements are
\begin{align}
\begin{tabular}{|lll|c|}
\hline
 $q$ & $m_1$ & $m_2$ & $\Mae{P \to}{a_{qm_1}^\dagger a_{qm_2}}{P \to}$ \\ \hline
 $u$ & $\up$ & $\down$ & 2/3 \\
 $u$ & $\down$ & $\up$ & 2/3 \\
 $d$ & $\up$ & $\down$ & -1/6 \\ 
 $d$ & $\down$ & $\up$ & -1/6 \\ \hline
\end{tabular}
\nn\end{align}

The dPDFs we consider are invariant under spin flip (they are only sensitive to diparton spin correlations), so we can simply use a spin-up proton. 
The dPDF for $dd$ in all spin combinations vanishes in the bag model since there is only one valance $d$ quark in the proton. The nonvanishing matrix elements are
\begin{align}
\begin{tabular}{|llllll|c|}
\hline
 $q_1$ & $q_2$ & $m_1$ & $m_2$ & $m_3$ & $m_4$ & $\Mae{P \up}{a_{q_1m_1}^\dagger a_{q_2 m_2}^\dagger a_{q_2 m_4} a_{q_1m_3}}{P \up}$ \\ \hline
 u & u & $\up$ & $\up$ & $\up$ & $\up$ & 4/3 \\
 u & u & $\up$ & $\down$ & $\up$ & $\down$ & 1/3  \\
 u & u & $\down$ & $\up$ & $\down$ & $\up$ & 1/3 \\ 
 u & u & $\up$ & $\down$ & $\down$ & $\up$ & 1/3 \\
 u & u & $\down$ & $\up$ & $\up$ & $\down$ & 1/3 \\ \hline
 u & d & $\up$ & $\up$ & $\up$ & $\up$ & 1/3 \\
 u & d & $\up$ & $\down$ & $\up$ & $\down$ & 4/3 \\
 u & d & $\down$ & $\up$ & $\down$ & $\up$ & 1/3 \\ 
 u & d & $\up$ & $\down$ & $\down$ & $\up$ & -2/3 \\
 u & d & $\down$ & $\up$ & $\up$ & $\down$ & -2/3 \\ \hline 
\end{tabular}
\nn\end{align}
Note that due to these spin-flavor correlations, the dPDF for $uu$ and $ud$  do not simply differ by an overall factor, as is the case for the single PDF.

\subsection{Rigid Bag}
\label{subsec:jaffe}

\begin{figure}[t!]
\centering
\includegraphics[width=0.4\textwidth]{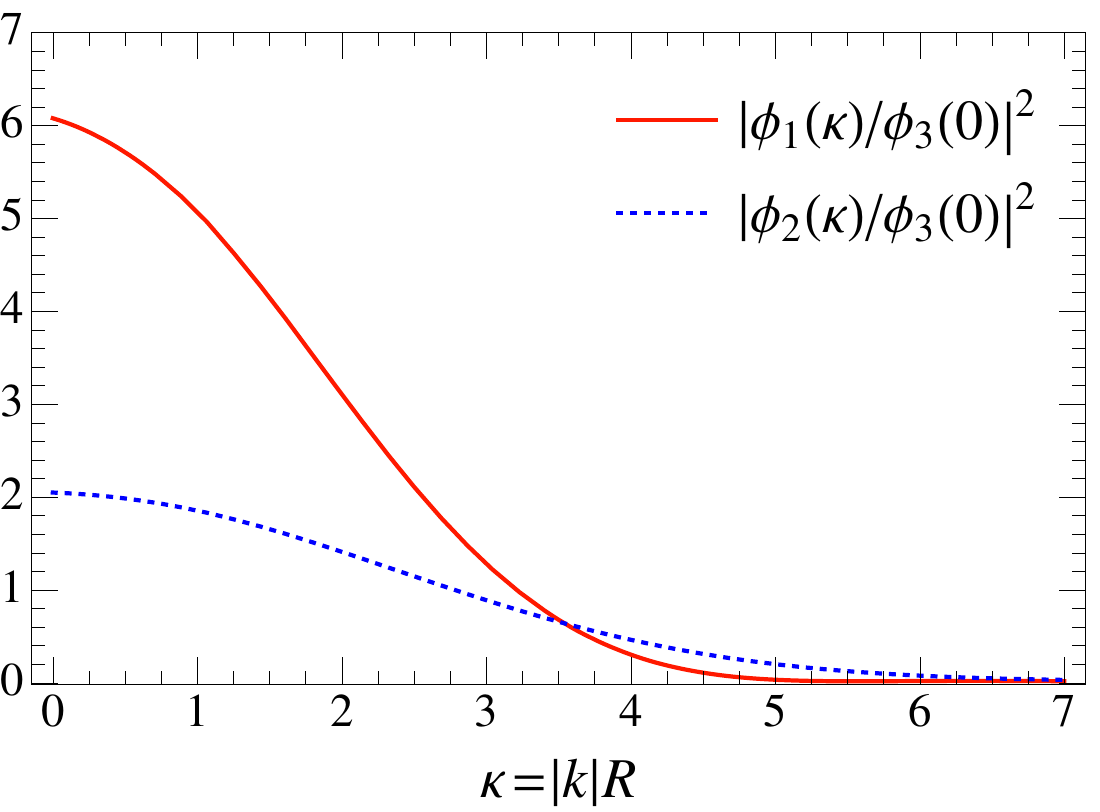}
\caption{Plot of the PY factors which enter the  calculation  of the single PDF (dotted blue) and double PDF (solid red). They suppress the PDFs in the unphysical regions $x>1$ and $x<0$.}
\label{fig:phisq}
\end{figure}

\begin{figure}[t!]
\centering
\includegraphics[width=0.43\textwidth]{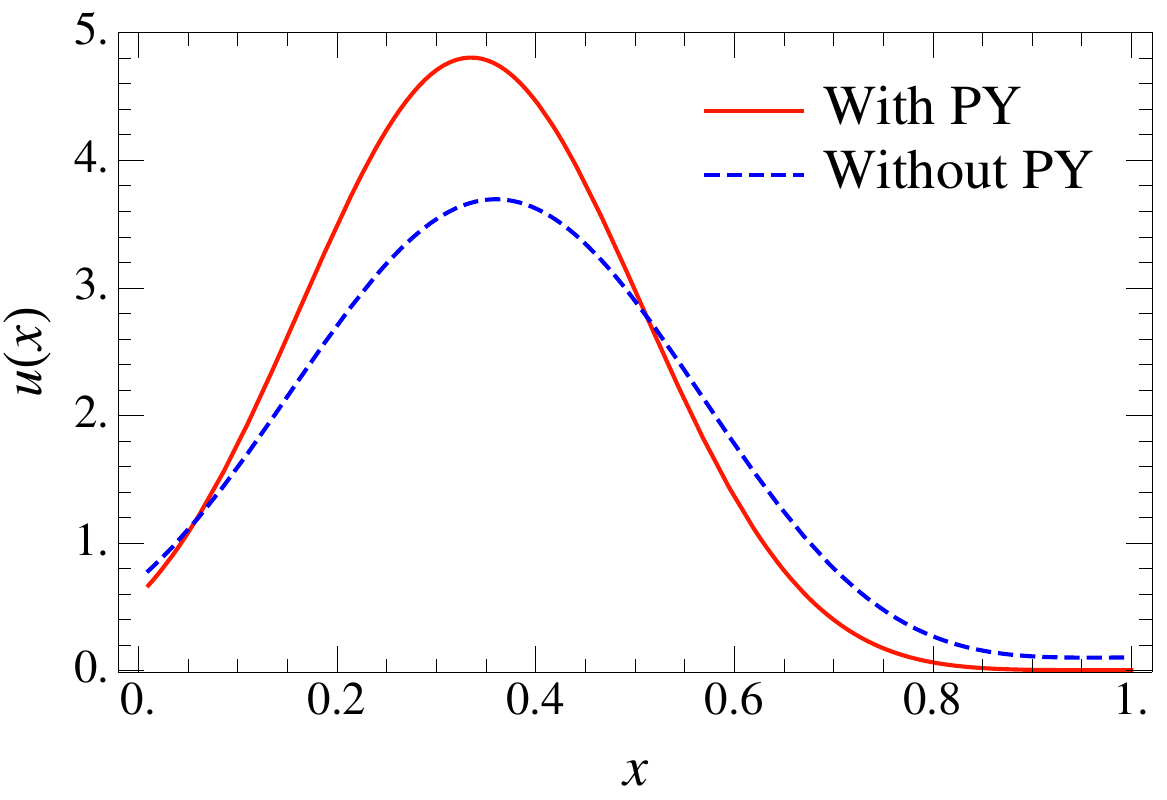}
\caption{Plot of the bag model proton PDF $\spdfa{u}(x)$ with (solid red) and without (dotted blue) PY factors.}
\label{fig:spdf}
\end{figure}

For a rigid bag, the overlap of empty bag states is
\begin{align} \label{eq:rigid}
\braket{0,{\bf r=a}}{0,{\bf r=b}}&=\de^3({\bf a-b})
\,.\end{align}
The only change to the single PDF in  \eq{fq} is that it removes the PY factor $|\phi_2({\bf k})|^2/|\phi_3({\bf 0})|^2$. This factor suppresses the ``leakage" of the PDF into the unphysical regions $x<0$ and $x>1$, without affecting the integral over all $x$, see \subsec{norm}. The PY factor is plotted in \fig{phisq}, and the PDF with and without the PY factor is shown in \fig{spdf}. 


Similarly, the rigid bag overlap in \eq{rigid} removes the factor $|\phi_1({\bf k})|^2/|\phi_3({\bf 0})|^2$ (also plotted in \fig{phisq}) from the double PDF in \eq{dPDF1b}. In this case, the dPDF factors, and there are no correlations between the momentum fractions $x_1$ and $x_2$, which is a clear shortcoming of treating the bag as rigid. At $\bf k_\perp =0$, the rigid bag dPDF takes a particularly simple form:
\begin{align} \label{eq:factor}
 \dpdf{q_1q_2}(x_1,x_2,{\bf k_\perp = \bf 0}) = \frac{c_{q_1q_2}}{n_{q_1} n_{q_2}}\, \spdfa{q_1}(x_1) \spdfa{q_2}(x_2)
\,,\end{align}
where the coefficient $c_{q_1q_2}$ is fixed by the spin-flavor wave function
\begin{align} \label{eq:cqq}
 c_{q_1q_2} =& 
\sum_{m_1=m_3 \atop m_2=m_4}\!\! \Mae{P,{\bf r = 0}}{a_{q_1m_1}^\dagger({\bf 0}) a_{q_2 m_2}^\dagger({\bf 0}) \nn \\
& \times a_{q_2 m_4}({\bf 0}) a_{q_1m_3}({\bf 0})}{P,{\bf r = 0}}
\,.\end{align}
From the tables in Sec.~\ref{subsec:spin_cor}, we find that $c_{uu}=c_{ud}=2$.

\begin{figure*}
\centering
\includegraphics[width=0.45\textwidth]{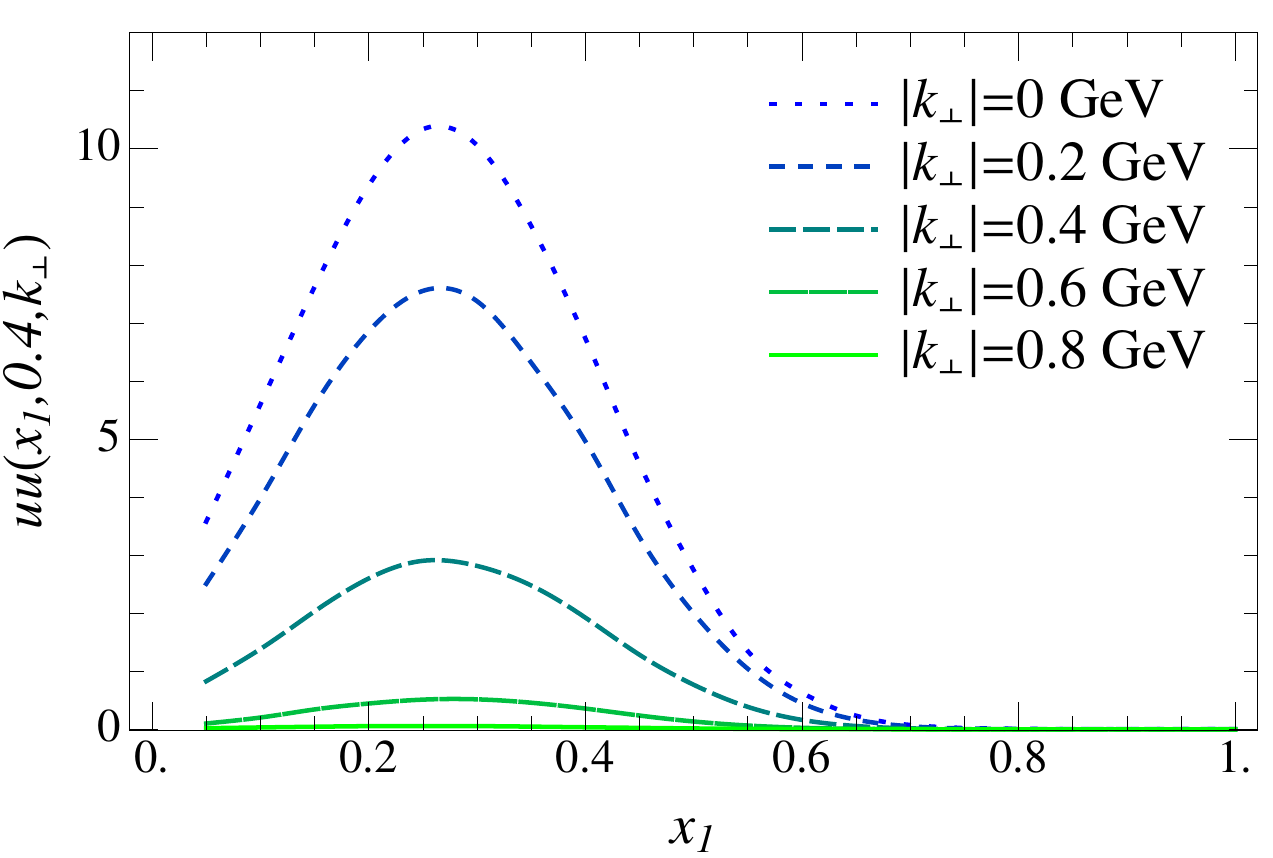} \hspace{1.25cm}
\includegraphics[width=0.45\textwidth]{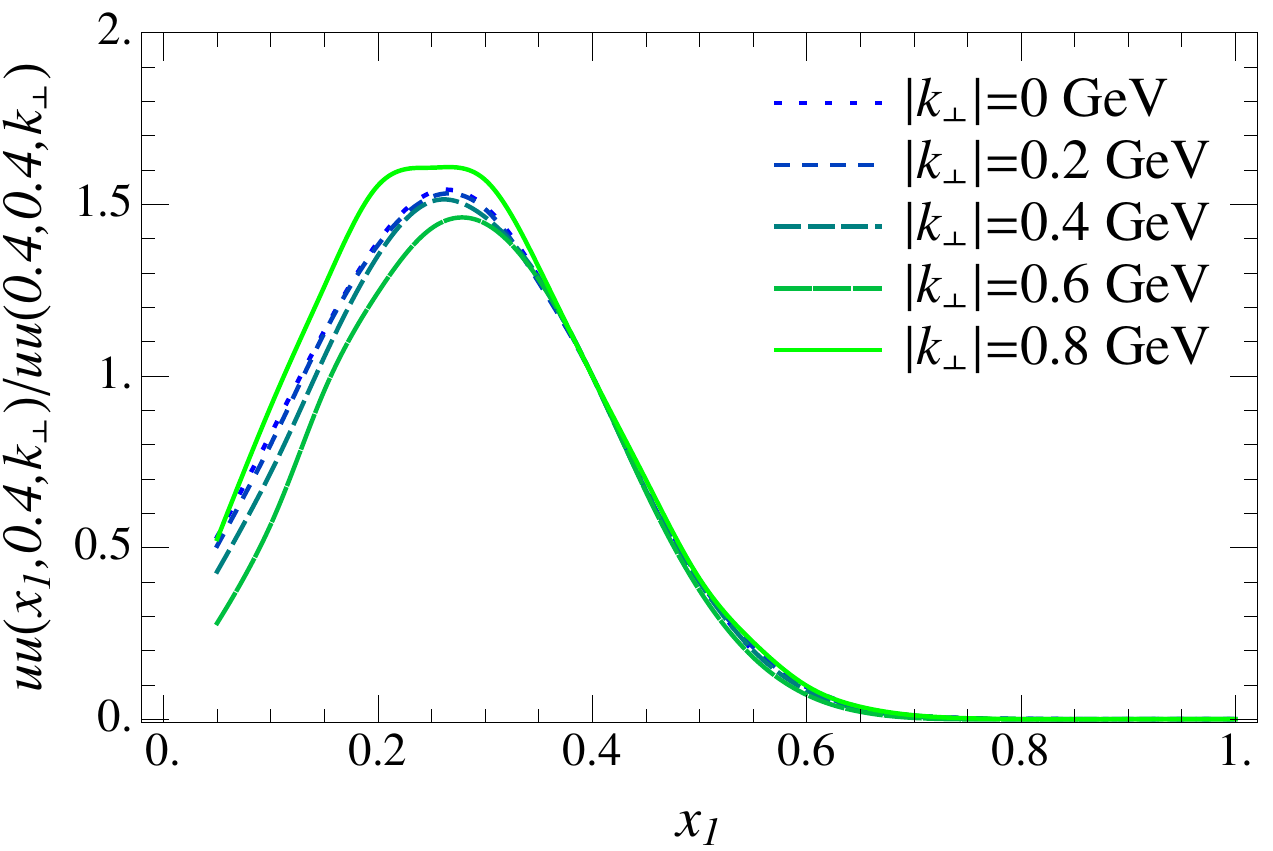}
\caption{The double PDF $\dpdfa{uu}(x_1, x_2,{\bf k_\perp})$ as a function of $x_1$ and $|\bf k_\perp|$ for fixed $x_2=0.4$. The right panel tests the ansatz in \eq{zfact} that $x_i$ and ${\bf k_\perp}$ are uncorrelated. This holds reasonably well, since the different $|\bf k_\perp|$ curves are nearly identical.}
\label{fig:fx1kp}
\end{figure*}
\begin{figure*}
\centering
\includegraphics[width=0.45\textwidth]{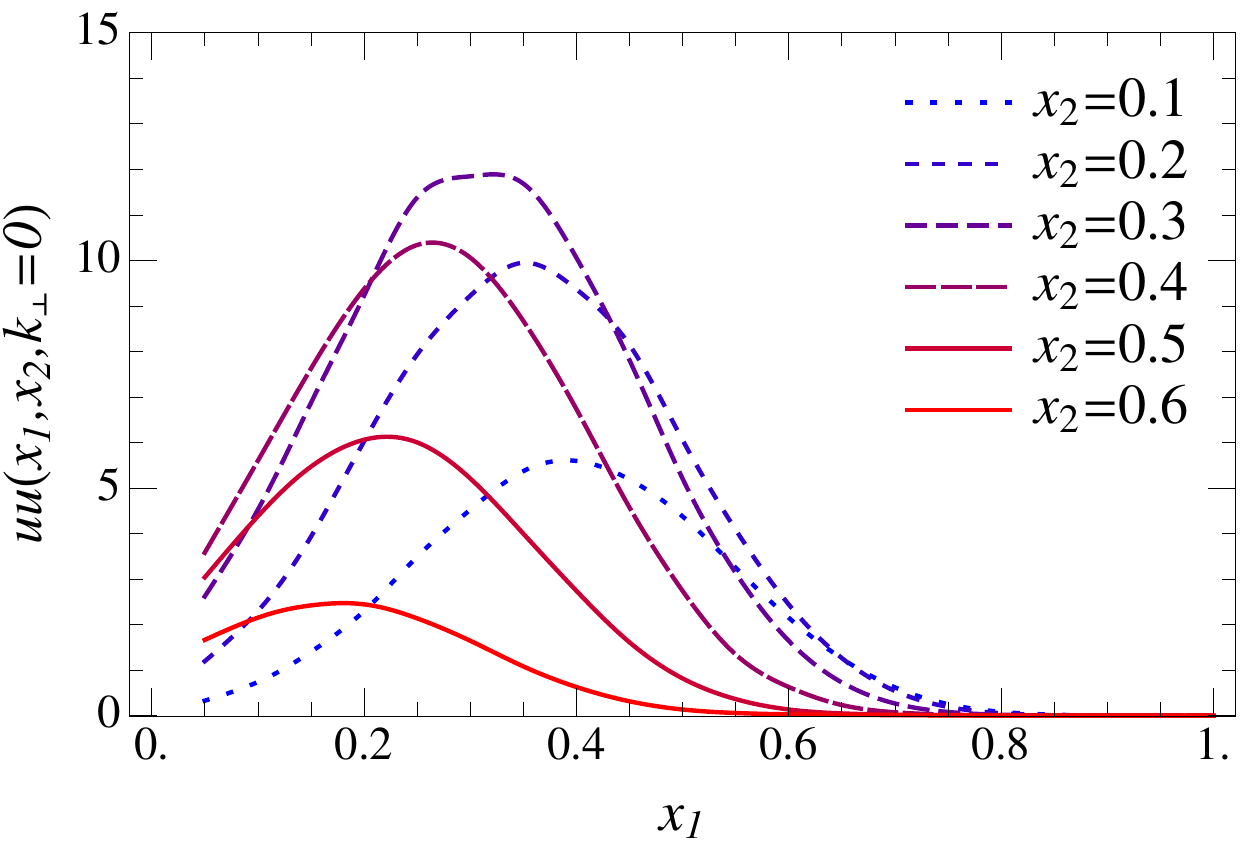} \hspace{1.25cm}
\includegraphics[width=0.44\textwidth]{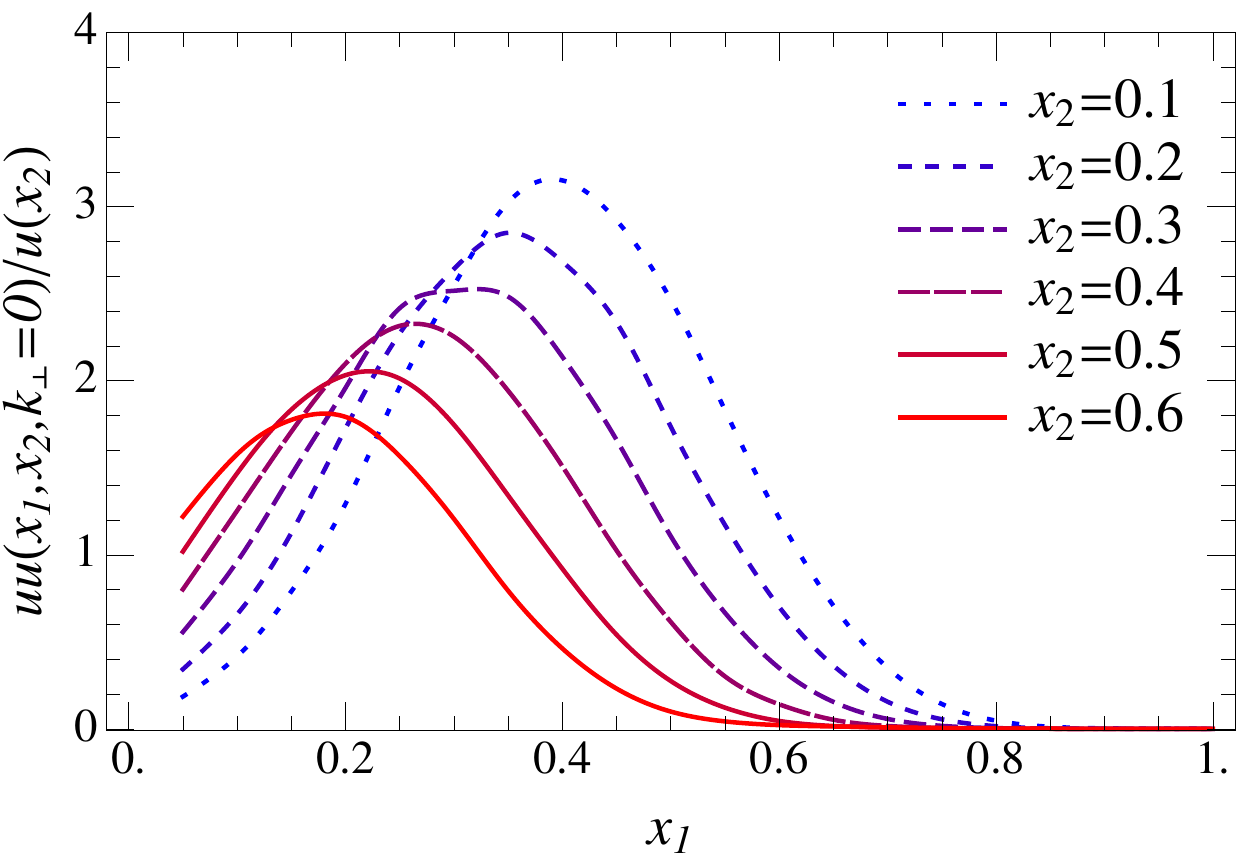}
\caption{The double PDF $\dpdfa{uu}(x_1, x_2,{\bf k_\perp})$ as a function of $x_1$ and $x_2$ for fixed ${\bf k_\perp=0}$. In the right panel, we divide by $\spdfa{u}(x_2)$ to test the often-used assumption in \eq{xfact} that the $x_i$ are uncorrelated. This clearly fails, since the ratio depends strongly on $x_2$.}
\label{fig:fx1x2}
\end{figure*}
%
%
%

\subsection{Normalization}
\label{subsec:norm}

The normalization of the single PDF and dPDF is given by integrating over all $x$, including unphysical regions. Both treatments of the bag in \eq{bag_overlap} will be considered. The single PDF in a rigid bag gives
\begin{align} \label{eq:norm}
\int\!\df x\, \spdfa{q}(x)
 & = n_q \int\!\df x\, \frac{2M}{(2\pi)^2}\, \!\int_{|xM - \W/R|}^\infty\!\!\!\!\!\!\!\!\! \df |{\bf k}|\, |{\bf k}|\, \bar{\widetilde{\Psi}}_m({\bf k}) \frac{\bnslash}{2} \widetilde{\Psi}_m({\bf k}) 
 \nn \\ &
 = 2n_q \int \! \frac{\df^3 {\bf k}}{(2\pi)^3}\, \widetilde{\Psi}^\dagger({\bf k}) \frac{\nslash \bnslash}{4} \widetilde{\Psi}({\bf k}) 
 \nn \\ &
  = n_q \int \! \df^3 {\bf y}\, |\Psi({\bf y})|^2 
 \nn \\ &
  = n_q
\,.\end{align}
Here we used that
\begin{align}
  \ga^0 \frac{\bnslash}{2} = \frac{\nslash \bnslash}{4}
 \,, \qquad
  \frac{\nslash \bnslash}{4}+\frac{\bnslash \nslash}{4} =1
\,.\end{align}
This second equation and the symmetry between $n$ and $\bn$ implies that we could replace $\nslash \bnslash/4 \to 1/2$ in \eq{norm}. The corresponding calculation with a flexible bag, i.e.\ including the PY factor, is
\begin{align} \label{eq:norm2}
&\int\!\df x\, \spdfa{q}(x)
\nn \\ & \quad
 = \frac{2M n_q}{(2\pi)^2}  \int\!\df x  \int_{|xM - \W/R|}^\infty\!\!\!\!\! \df |{\bf k}|\, |{\bf k}|\, \bar{\widetilde{\Psi}}_m({\bf k}) \frac{\bnslash}{2} \widetilde{\Psi}_m({\bf k}) \frac{|\phi_2({\bf k})|^2}{|\phi_3(0)|^2} 
\nn \\ & \quad
 = \frac{2n_q}{|\phi_3(0)|^2} \int\!  \frac{\df^3 {\bf k}}{(2\pi)^3}\,
\widetilde{\Psi}_m^\dagger({\bf k}) \frac{\nslash \bnslash}{4} \widetilde{\Psi}_m({\bf k})\, |\phi_2({\bf k})|^2 
\nn \\ & \quad
= \frac{n_q}{|\phi_3(0)|^2} \!\int\!  \frac{\df^3 {\bf k}}{(2\pi)^3}
 \!\int\! \df^3 {\bf x}_1\, \df^3 {\bf y}_1\, e^{\img {\bf k} \cdot {\bf x}_1}\, \Psi^\dagger({\bf y}_1 \!-\! {\bf x}_1) \Psi({\bf y}_1)
 \nn \\ & \qquad \times
\int\! \df^3 {\bf x}_2\, e^{-\img {\bf k} \cdot {\bf x}_2} \Big[\int\! \df {\bf y}_2\,
 \Psi^\dagger({\bf y}_2-{\bf x}_2) \Psi({\bf y}_2) \Big]^2
\nn \\ & \quad
= n_q \,,
\end{align}
and has the same normalization. However, the PY factor reduces the PDF at unphysical $x$. Specifically, 2\% of the contribution to the integral in \eq{norm2} is from the unphysical region, compared to 11\% in \eq{norm}.

\begin{figure}[b]
\centering
\includegraphics[width=0.43\textwidth]{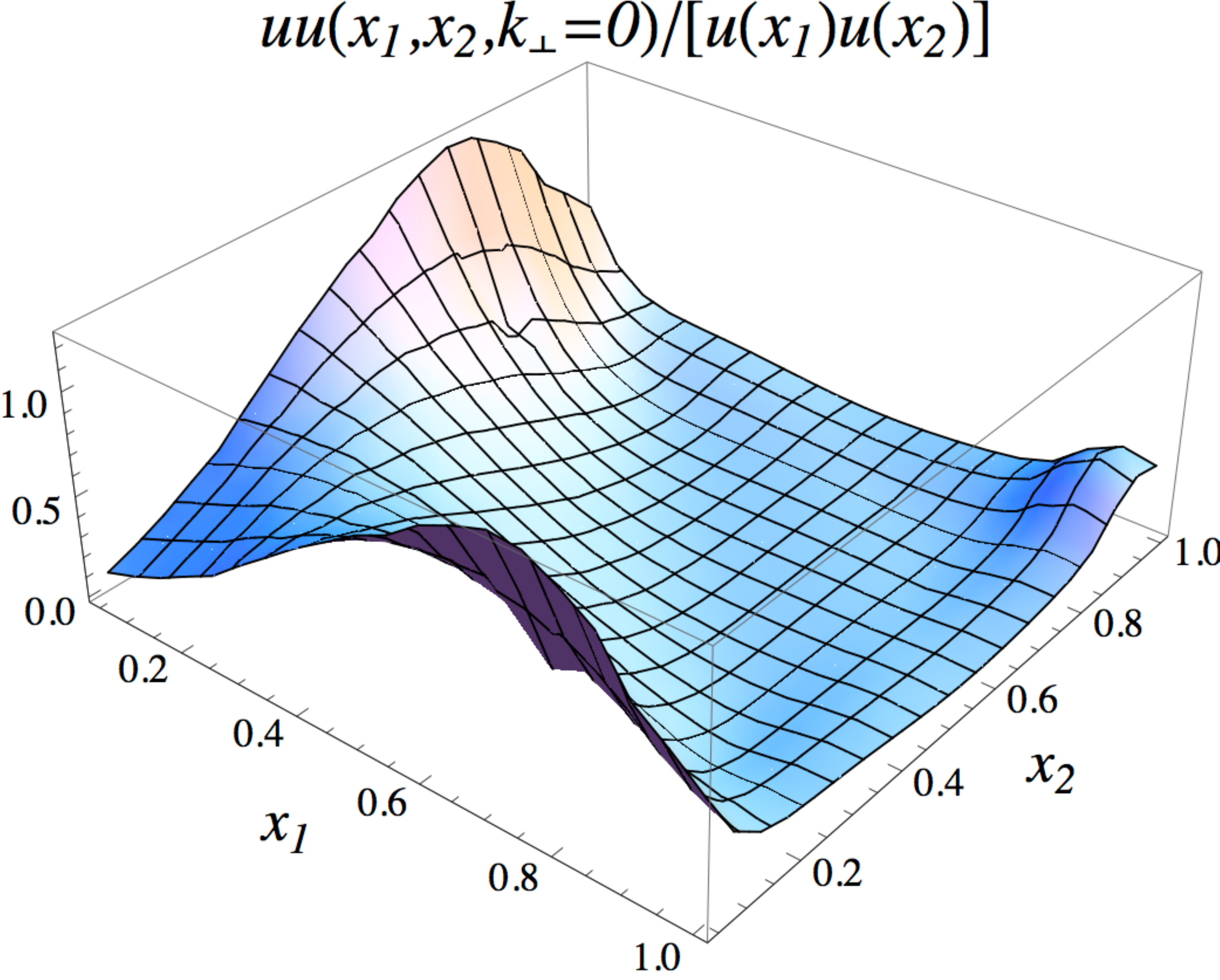}
\caption{The correlation between the momentum fractions of two $u$ quarks in the proton is shown by plotting the ratio of the double PDF $\dpdfa{uu}(x_1,x_2,\mathbf{k}_\perp=0)$ to the product of two single PDFs $\spdfa{u}(x_1) \spdfa{u}(x_2)$.}
\label{fig:3d}
\end{figure}

For the dPDF, the normalization for the rigid bag follows from \eqs{factor}{norm}
\begin{align} \label{eq:dpdf_norm}
  & \int\! \df x_1\, \df x_2\, \df {\bf z}_\perp \dpdf{q_1 q_2}(x_1,x_2,{\bf z}_\perp)
  \nn \\ & \qquad
  = \frac{c_{q_1q_2}}{n_{q_1} n_{q_2}} \int\! \df x_1\, \df x_2\, \spdfa{q_1}(x_1) \spdfa{q_2}(x_2) 
  \nn \\ & \qquad
  = c_{q_1q_2}
\,, \end{align}
where the coefficient $c_{q_1 q_2}$ is given in \eq{cqq}. The calculation including the PY factor is similar to \eq{norm2}
\begin{align}
& \int\!\df x_1\,\df x_2\, \df^2{\bf z_\perp}\dpdf{qq}(x_1,x_2,{\bf z}_\perp) 
\nn \\
& \qquad = 4 c_{q_1q_2}  \int\!  \frac{\df^3 {\bf k}_1}{(2\pi)^3}\,  \frac{\df^3 {\bf k}_2}{(2\pi)^3}\, 
 \widetilde{\Psi}^\dagger({\bf k}_1) \frac{\nslash \bnslash}{4} \widetilde{\Psi}({\bf k}_1)
 \nn \\ & \qquad \times
 \widetilde{\Psi}^\dagger({\bf k}_2) \frac{\nslash \bnslash}{4} \widetilde{\Psi}({\bf k}_2)\, \frac{|\phi_1({\bf k}_1 + {\bf k}_2)|^2}{|\phi_3(0)|^2}
\nn \\ & \qquad
= c_{q_1q_2} [1+\ord{<1\%}]
\,.\end{align}
The small correction with respect to \eq{dpdf_norm} arises because we can no longer replace $\nslash \bnslash/4 \to 1/2$. Specifically, \eq{Psi_mom} implies
\begin{align} \label{eq:psi_mom2}
   \widetilde{\Psi}^\dagger({\bf k}) \frac{\nslash \bnslash}{4} \widetilde{\Psi}({\bf k}) &= 
   \frac{\pi R^3 \W^2}{2(\W^2 - \sin^2 \W)}
 (s_1^2 + 2s_1 s_2 {\hat {\bf k}}_z + s_2^2) 
 \,,\nn \\
  \widetilde{\Psi}^\dagger({\bf k}) \widetilde{\Psi}({\bf k}) &= 
   \frac{\pi R^3 \W^2}{(\W^2 - \sin^2 \W)}
 (s_1^2 + s_2^2) 
\,\end{align}
Since the momenta ${\bf k}_{1z}$ and ${\bf k}_{2z}$ become correlated through $\phi_1({\bf k}_1 + {\bf k}_2)$, this implies that $\langle {\bf k}_{1z} {\bf k}_{2z} \rangle \neq \langle {\bf k}_{1z} \rangle \langle {\bf k}_{2z} \rangle = 0$.

\begin{figure*}
\centering
\includegraphics[width=0.45\textwidth]{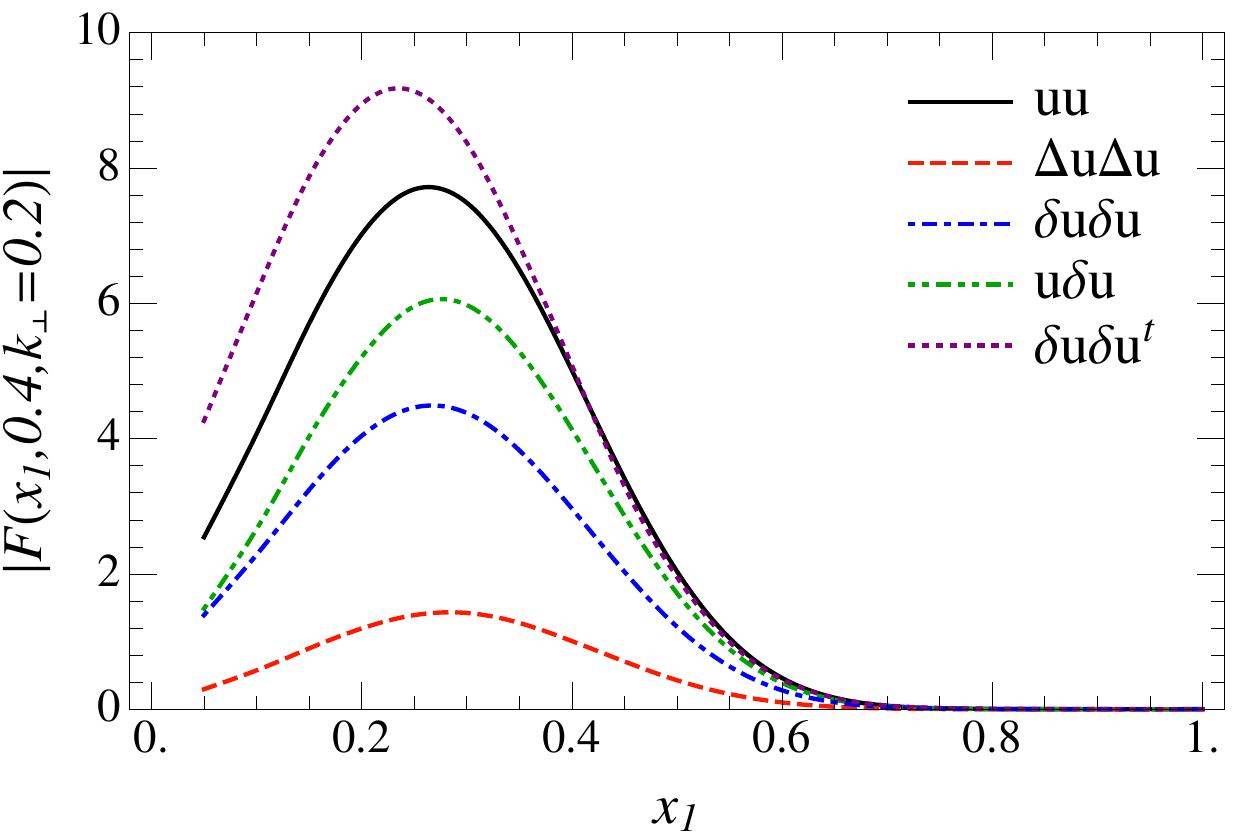}  \hspace{1.25cm}
\includegraphics[width=0.45\textwidth]{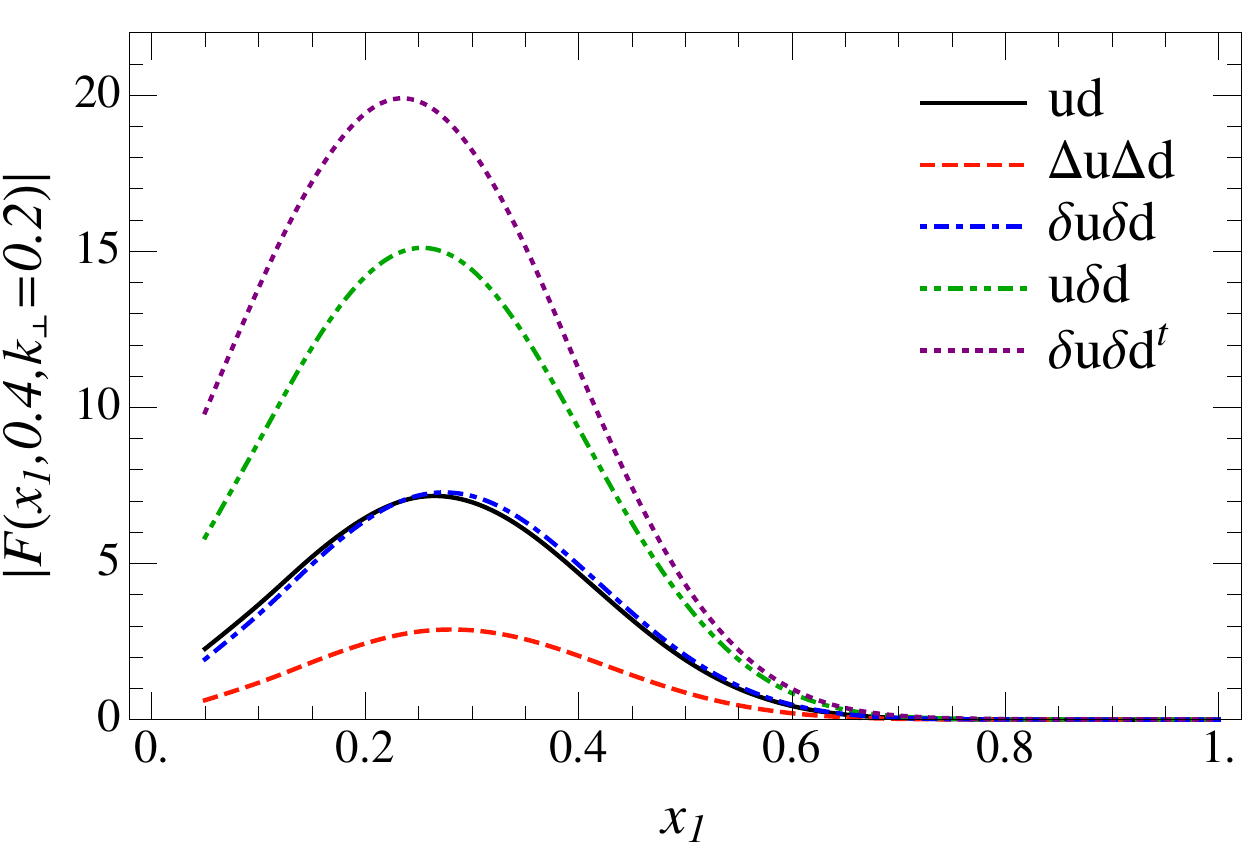} \\
\includegraphics[width=0.45\textwidth]{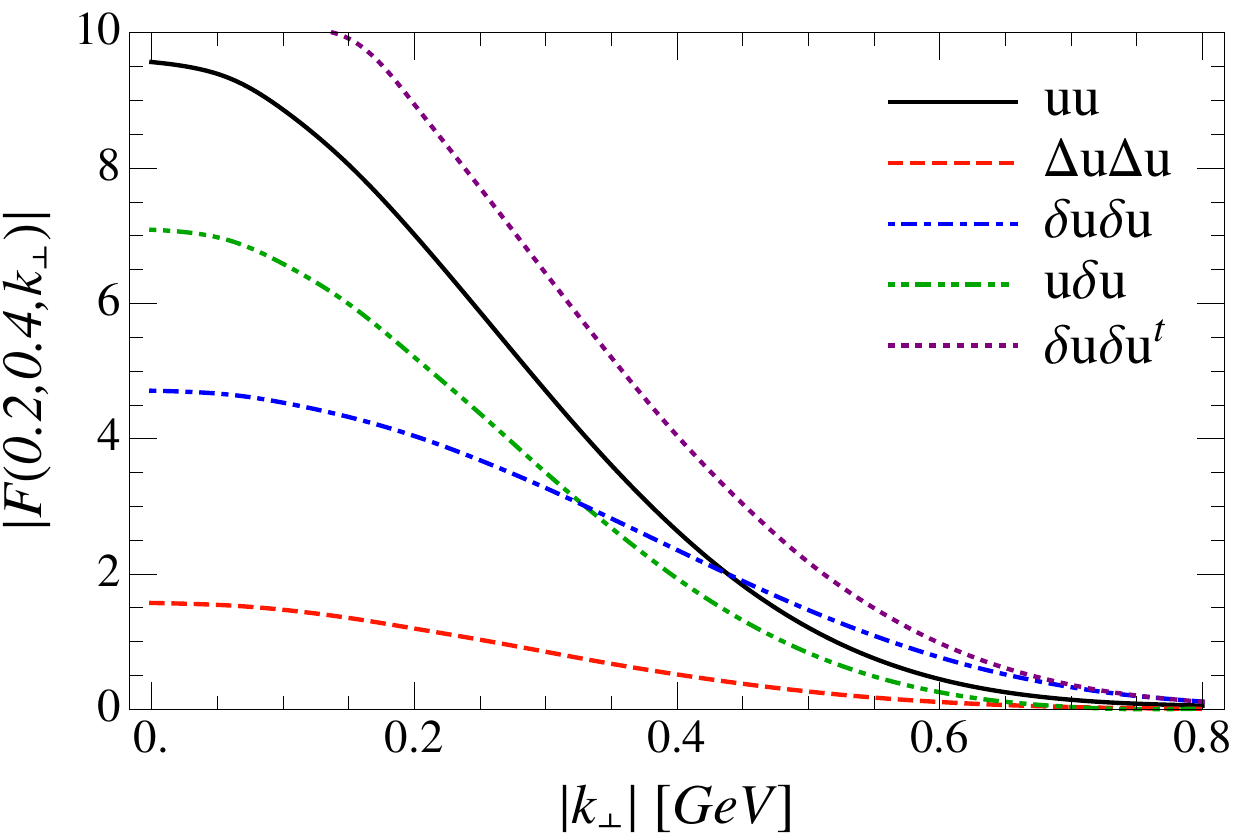}  \hspace{1.25cm}
\includegraphics[width=0.45\textwidth]{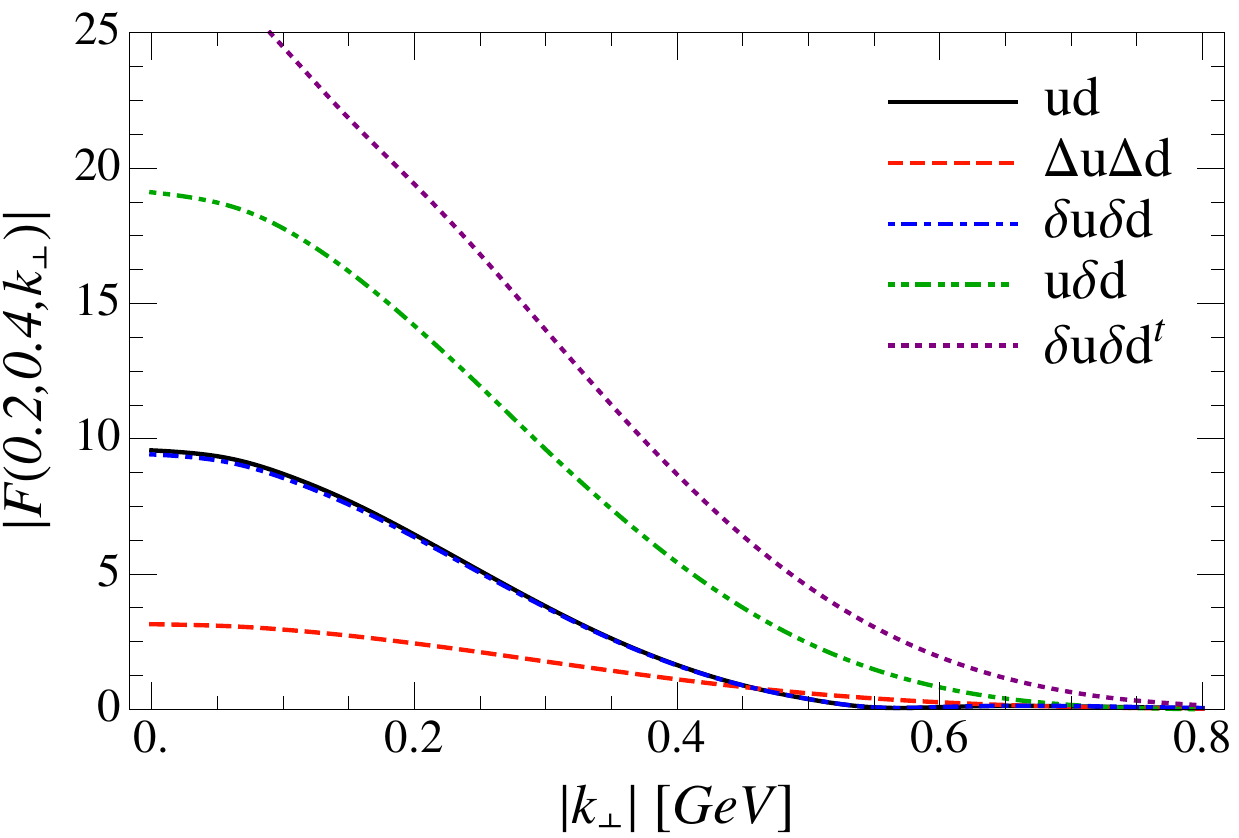}
\caption{Comparison of the double PDF spin structures as functions of $x_1$ or $\bf |k_\perp|$, keeping the other variables fixed. The left panels show the $uu$ double PDFs, and the right panels show the $ud$ double PDFs. The $u \de u$, $\de u \de u$, $\de u \de u^t$, $\Delta u \Delta d$, $u \de d$ and $\de u \de d^t$ distributions are negative, and we have changed their sign in these plots. Note that $ud$ and $\de u \de d$ are almost indistinguishable.}
\label{fig:fspin}
\end{figure*}
%
%
%

\section{Parton Correlations}
\label{sec:results}

We are now ready to investigate the size of the various diparton correlation effects using the bag model dPDFs. We start by studying the dependence of the dPDF $\dpdfa{uu}(x_1,x_2,{\bf k_\perp})$ on $x_1$ and $|{\bf k_\perp}|$, keeping $x_2=0.4$ fixed for simplicity. As the left panel of \fig{fx1kp} shows, the dPDF reduces significantly with increasing $|{\bf k_\perp}|$. In the right panel we test the ansatz in \eq{zfact} that the dependence on $x_i$ and $\bf k_\perp$ is uncorrelated, by dividing by $\dpdfa{uu}(0.4,0.4,{\bf k_\perp})$. If the ansatz holds, the universal transverse function $G({\bf k_\perp})$ should drop out in this ratio, making the result independent of ${\bf k_\perp}$. As the plot shows, this seems to holds quite well. It only breaks down for the largest values of $|{\bf k_\perp}|$, where the dPDF is orders of magnitude smaller than at $|{\bf k_\perp}|=0$. We also note that there is some leakage into the unphysical region $x_1+x_2>1$, as was the case for the single PDF in \fig{spdf}, though this effect is reasonably small. 

Next we explore the $x_1,x_2$ dependence of dPDF $\dpdfa{uu}(x_1,x_2,{\bf k_\perp})$ for ${\bf k_\perp=0}$, which is shown in \fig{fx1x2}. As $x_2$ is increased, the peak of the $x_1$ distribution moves to smaller $x_1$, responding to the reduced momentum available. The peak height reduces as well, though not for small $x_2$ since the bag model only describes the valence quarks. To test the factorization ansatz in \eq{xfact} for $n=0$, we divide by $\spdfa{u}(x_2)$ in the right panel. Since the resulting distributions clearly still depend on $x_2$, correlations between $x_1$ and $x_2$ are important. Inclusion of the factor of $(1-x_1-x_2)^n$ does not alter this conclusion. The correlations can also be seen in the three-dimensional plot of Fig.~\ref{fig:3d}. We  remind the reader that this conclusion depends on the treatment of the bag, since $x_1$ and $x_2$ would be uncorrelated if a rigid bag was assumed (see \subsec{jaffe}). 

The relative size of the various spin structures in \subsec{spin_cor} are studied in \fig{fspin}. They are shown as a function of $x_1$ (top row) and $|{\bf k_\perp}|$ (bottom row), keeping all other variables fixed. All spin structures show a similar dependence on $x_1$ and ${\bf k}_\perp$, though there is a hierarchy between their sizes. \fig{fspin} also illustrates the differences between the $uu$ (left column) and $ud$ (right column) dPDF.  Unlike the single PDF, where the difference between $\spdfa{u}$ and $\spdfa{d}$ was simply an overall factor of $n_u/n_d = 2$, the dPDF has more flavor dependence. This arises through the spin dependence and the correlations in the spin-flavor wave function. As \fig{fspin} shows, the difference between $uu$ and $ud$ is fairly small. However, the spin correlations are about twice as big for $ud$ than for $uu$. 

The shape of the $|{\bf k_\perp}|$ dependence is reasonably well described by a Gaussian,
\begin{align} \label{eq:gauss}
G({\bf k_\perp}) \approx \frac{1}{2\pi \si^2} e^{- {\bf k}_\perp^2/(2 \si^2)}
\,.\end{align}
The width $\si$ depends slightly on the spin structure:
\begin{align}
&\begin{tabular}{|c|ccccc|}
\hline
 $$ & $uu$ & $\De u \De u$ & $\de u \de u$ & $u \de u$ & $\de u \de u^t$ \\ \hline
 $\si$ (GeV) & 0.25 & 0.27 & 0.32 & 0.25 & 0.29 \\ \hline
\end{tabular}
\nn \\[1ex]
&\begin{tabular}{|c|ccccc|}
\hline
 $$ & $ud$ & $\De u \De d$ & $\de u \de d$ & $u \de d$ & $\de u \de d^t$ \\ \hline
 $\si$ (GeV) & 0.22 & 0.27 & 0.22 & 0.25 & 0.26 \\ \hline
\end{tabular}
\nn\end{align}
Note that in the bag model $u \de d = d \de u$. 

\section{Conclusions}
\label{sec:conclusions}

We have computed the dPDFs using a bag model for the proton. The bag model results should be treated as the dPDFs at a low scale, which can then be evolved to higher energy using the known QCD evolution equations~\cite{Manohar:2012jr,Manohar:DPS2}. We find substantial diparton correlations in the proton in spin, flavor, and momentum fraction, which have traditionally been ignored in analyses of double parton scattering, but only a small correlation with the transverse momentum $\mathbf{k}_\perp$. The $uu$ and $ud$ dPDFs are not simply related to each other, or to the single PDFs $u$ and $d$, because of the spin-flavor correlations in the proton quark model wave function in \eq{ketp}. The results in this paper provide quantitative results for these diparton correlations, which will help in the experimental analysis of double parton scattering at the LHC.

\acknowledgments

We would like to thank G.~A.~Miller and R.~Jaffe for helpful discussions.
This work is supported in part by DOE Grant No.~DE-FG02-90ER40546. 

\appendix

\section{}
\label{app:integrals}

We collect simplified expressions for the bag model wave function in momentum space and the functions $\phi_n$ needed for the PY projection. Several of these results were already obtained in Ref.~\cite{Schreiber:1991tc}. The Fourier transform of the wave function is
\begin{align} \label{eq:Psi_mom}
 \widetilde{\Psi}_m({\bf k}) &= \int\! \df^3 {\bf x}\, e^{\img {\bf k} \cdot {\bf x}}\,    \Psi_m({\bf x})
 \nn \\ &
 = \frac{2\pi \W R^{3/2}}{\sqrt{\W^2 - \sin^2 \W}}
 \begin{pmatrix}
  s_1(\kappa) \chi_m \\
  s_2(\kappa)\, {\bf \hat k} \sdt {\bm \si} \, \chi_m
 \end{pmatrix}
\,,\end{align}
where $\kappa= |{\bf k}|R$ and
\begin{align}
  s_1(\kappa) &= \frac{1}{\kappa} \Big[\frac{\sin (\kappa - \W)}{\kappa- \W} - \frac{\sin (\kappa + \W)}{\kappa + \W}\Big]
  \,, \nn \\ 
  s_2(\kappa) &= 2j_0(\W) j_1(\kappa) - \frac{\kappa}{\W} s_1(\kappa)
\,,\end{align}
and $\chi_m$ is defined in \eq{chidef}.
For the unpolarized and longitudinally polarized single PDFs this leads to
\begin{align}
 \bar{\widetilde{\Psi}}_m \frac{\bnslash}{2} \widetilde{\Psi}_m &=
\frac{\pi R^3\W^2}{2(\W^2 - \sin^2 \W)}\,
 (s_1^2 + s_2^2 + 2s_1 s_2 \hat k_z)
\,, \nn \\ 
 \bar{\widetilde{\Psi}}_m \frac{\bnslash}{2} \ga_5 \widetilde{\Psi}_m  &= (-1)^{m+3/2}
\frac{\pi R^3 \Omega^2}{2(\W^2 - \sin^2 \W)}
\nn \\ & \quad \times 
[s_1^2 \!+\! s_2^2 (1-2 {\bf \hat k_\perp}^2) \!+\! 2s_1 s_2 \hat k_z] 
\,.\end{align}
For the transversely polarized PDF we need
\begin{align}
 & \bar{\widetilde{\Psi}}_\up \frac{\bnslash}{2} \ga_\perp^1 \ga_5 \widetilde{\Psi}_\down
 + \bar{\widetilde{\Psi}}_\down \frac{\bnslash}{2} \ga_\perp^1 \ga_5 \widetilde{\Psi}_\up
 \nn \\
 & \quad
 = \frac{\pi R^3 \Omega^2}{\W^2 - \sin^2 \W}\,
[s_1^2 \!+\! s_2^2 (1-2{\hat k_x}^2) \!+\! 2s_1 s_2 \hat k_z] 
\,.\end{align}

The functions $\phi_n$, used in the PY projection, are
\begin{align} 
 |\phi_n({\bf p})|^2 &= \frac{2^{4-n}\pi R^3\W^{n-2}}{\kappa(\W^2 - \sin^2 \W)^n} \int_0^\W\! \frac{\df v}{v^{n-1}}\,  \sin \frac{2\kappa v}{\W}\, T^n(v)
\,,\end{align}
with
\begin{align}
T(v) &= \Big(\W \!-\! \frac{1-\cos 2\W}{2\W} \!-\! v \Big) \sin 2v
 - \Big(\frac{1}{2} \!+\! \frac{\sin 2\W}{2\W}\Big) \cos 2v 
\nn \\ & \quad
 + \frac{1}{2} + \frac{\sin 2\W}{2\W} - \frac{1- \cos 2\W}{2\W^2} v^2 
\,.\end{align}

\newpage

\section{}
\label{app:spins}

The relationship between the dPDFs $\dpdfk{}$ and $F$ defined in \subsec{spin_cor} is
\begin{align}
  \dpdfk{q_1 \de q_2}(x_1,x_2,{\bf k}_\perp) &= -\frac{\img M^2}{{\bf k}_\perp^2} \int\! \df {\bf z_\perp} e^{\img {\bf k}_\perp \cdot {\bf z}_\perp} ({\bf k}_\perp \sdt {\bf z}_\perp) 
  \nn \\ & \quad \times
  F_{q_1 \de q_2}(x_1,x_2,{\bf z}_\perp)
  \,, \nn \\
  \dpdfk{\De q_1 \de q_2}(x_1,x_2,{\bf k}_\perp) &= -\frac{\img M^2}{{\bf k}_\perp^2} \int\! \df {\bf z_\perp} e^{\img {\bf k}_\perp \cdot {\bf z}_\perp} ({\bf k}_\perp \sdt {\bf z}_\perp) 
  \nn \\ & \quad \times
  F_{\De q_1 \de q_2}(x_1,x_2,{\bf z}_\perp)
  \,, \nn \\
  \dpdfk{\de q_1 \de q_2}^t(x_1,x_2,{\bf k}_\perp) &= \frac{M^4}{|{\bf k}_\perp|^4} \int\! \df {\bf z_\perp} e^{\img {\bf k}_\perp \cdot {\bf z}_\perp} [2({\bf k}_\perp \sdt {\bf z}_\perp)^2\!-\!{\bf k}_\perp^2 {\bf z}_\perp^2] 
  \nn \\ & \quad \times  
  F_{\de q_1 \de q_2}^t(x_1,x_2,{\bf z}_\perp)
\,.\end{align}
The factors of ${\bf k} \sdt {\bf z}$ arise because $q_1 \de q_2$ and $\De q_1 \de q_2$ have $\perp$ angular momentum one, and $\de q_1 \de q_2^t$ has $\perp$ angular momentum two.
The other spin structures are not affected when switching to momentum space, so
 $F_{q_1q_2}(x_1,x_2,{\bf k}_\perp)$ is the Fourier transform of $F_{q_1q_2}(x_1,x_2,{\bf z}_\perp)$, etc.

\bibliography{dps}

\end{document}